\documentclass{cpcauth}
\usepackage{psfig}

\begin{document}
\begin{frontmatter}
\title{Phase transitions at surfaces, edges, and corners}

\author[ER,NA]{Michel Pleimling\thanksref{ca}}
\address[ER]{Institut f\"ur Theoretische Physik 1, Universit\"at Erlangen-N\"urnberg,
D--91058 
Erlangen, Germany}
\address[NA]{Laboratoire de Physique des Mat\'eriaux$^{\, 2}$, Universit\'e Henri
Poincar\'e Nancy I, B.P. 239,
F--54506 Vand{\oe}uvre l\`es Nancy Cedex, France}
\thanks[ca]{
e-mail: pleim@theorie1.physik.uni-erlangen.de}
\thanks[an]{Laboratoire associ\'e au CNRS UMR 7556}

\begin{abstract}
Results of large-scale Monte Carlo simulations of three-dimensional
Ising models with edges and corners are reviewed. At the ordinary transition,
angle dependent critical exponents are observed, whereas at the surface transition
edge and corner critical exponents are non-universal and depend on the details of the model.
The results obtained at the surface transition are compared to exact findings on critical
two-dimensional Ising models with different types of defect lines.
\end{abstract}
\end{frontmatter}
\section{Introduction}
Critical phenomena at perfect, flat surfaces of three-dimensional systems have been
the subject of intensive research during the last three decades \cite{Bin83,Die86,Dos92}.
For the three-dimensional semi-infinite Ising model two different ferromagnetic couplings
are usually introduced, depending on whether the neighbouring spins are both located 
at the surface, $J_s > 0$, or not, $J_b > 0$. The resulting phase diagram is well 
established. If the ratio of the surface coupling $J_s$ to the bulk
coupling $J_b$, $r=J_s/J_b$, is sufficiently small, the system undergoes at the bulk critical temperature
$T_{c}$ an ordinary transition, with the bulk and surface ordering occurring at the same temperature.
Beyond a critical ratio, $r > r_{sp} \approx 1.50$ \cite{Bin84,Rug93}, the surface orders at a higher
temperature $T_s > T_{c}$ at the surface transition, followed by the extraordinary transition
of the bulk at $T_{c}$. At the critical ratio $r_{sp}$, one encounters the multicritical
special transition point. 

The semi-infinite Ising model with a flat surface may be considered to be a special
case of a more complex wedge geometry where two planes meeting at an angle $\theta$
form an infinite edge. For $\theta= \pi$ the flat surface is recovered. Cardy \cite{Car83}
showed that at the ordinary transition edge critical exponents depending 
continuously on $\theta$ arise on purely geometrical grounds. For a given opening angle
$\theta$, however, the values of the critical exponents are expected to be universal and independent
on microscopic details like the strengths of the coupling constants or the lattice type.
Whereas edge singularities at the ordinary transition have been studied intensively,
especially in two dimensions \cite{Igl93}, edge critical 
behaviour at the surface transition of three-dimensional Ising models has been largely overlooked.
At the surface transition the critical fluctuations are essentially of two-dimensional
character. One may then argue that the edge should act like a 
local perturbation in a two-dimensional system.
In analogy with exact results obtained for two-dimensional
Ising models with defect lines \cite{Bar79}, intriguing nonuniversal edge critical behaviour
is therefore expected.
Similarly, corner critical exponents depending on the details of the model should also be observed
at the surface transition.

In this contribution, I discuss results on edge and corner critical behaviour obtained in large-scale
Monte Carlo simulations of three-dimensional Ising models \cite{Ple98,Ple99,Ple00}.
After presenting some details of the simulations in the next Section, I then discuss 
edge and corner criticality at the ordinary (Section 3) and at the surface transition
(Section 4). 
A brief summary concludes the paper.

\section{Models with edges and corners}
To introduce edges in Ising magnets defined on simple cubic lattices, 
periodic boundary conditions along one axis,
the $z$-axis, are applied. The remaining four free surfaces of the crystal may be
oriented in various ways leading to different opening angles $\theta$ at the edges.
As shown in Figure 1, pairs of (100) and (010) surfaces lead to four equivalent edges
with opening angles $\theta= \pi/2$. The intersection of (100) and (110) surfaces form
two pairs of edges with $\theta=\pi/4$ and $\theta= 3 \pi/4$, see Figure 2.
Besides the bulk coupling $J_b$ and the surface coupling $J_s$, further couplings may
be introduced \cite{Ple98,Ple99}: two neighbouring edge spins interact with the edge coupling
$J_e$, whereas an edge spin is coupled to its neighbouring surface spin by the edge-surface
interaction $J_{es}$.
The Hamiltonian is then given by
\begin{eqnarray}
H & = & -\sum\limits_{bulk} J_b S_{{\bf r}} S_{{\bf r}'}
 -\sum\limits_{surface} J_s S_{{\bf r}} S_{{\bf r}'} \nonumber \\
& & -\sum\limits_{edge-surface} J_{es} S_{{\bf r}} S_{{\bf r}'}
 -\sum\limits_{edge} J_e S_{{\bf r}} S_{{\bf r}'}
\end{eqnarray}
with spins $S_{{\bf r}}= \pm 1$ at sites ${\bf r}=(xyz)$.
For the study of corners we use free boundary conditions in all three directions.
The couplings between corner spins and neighbouring
edge spins are set equal to $J_e$, thus regarding corners as endpoints of edges.

\begin{figure}
\centerline{\psfig{figure=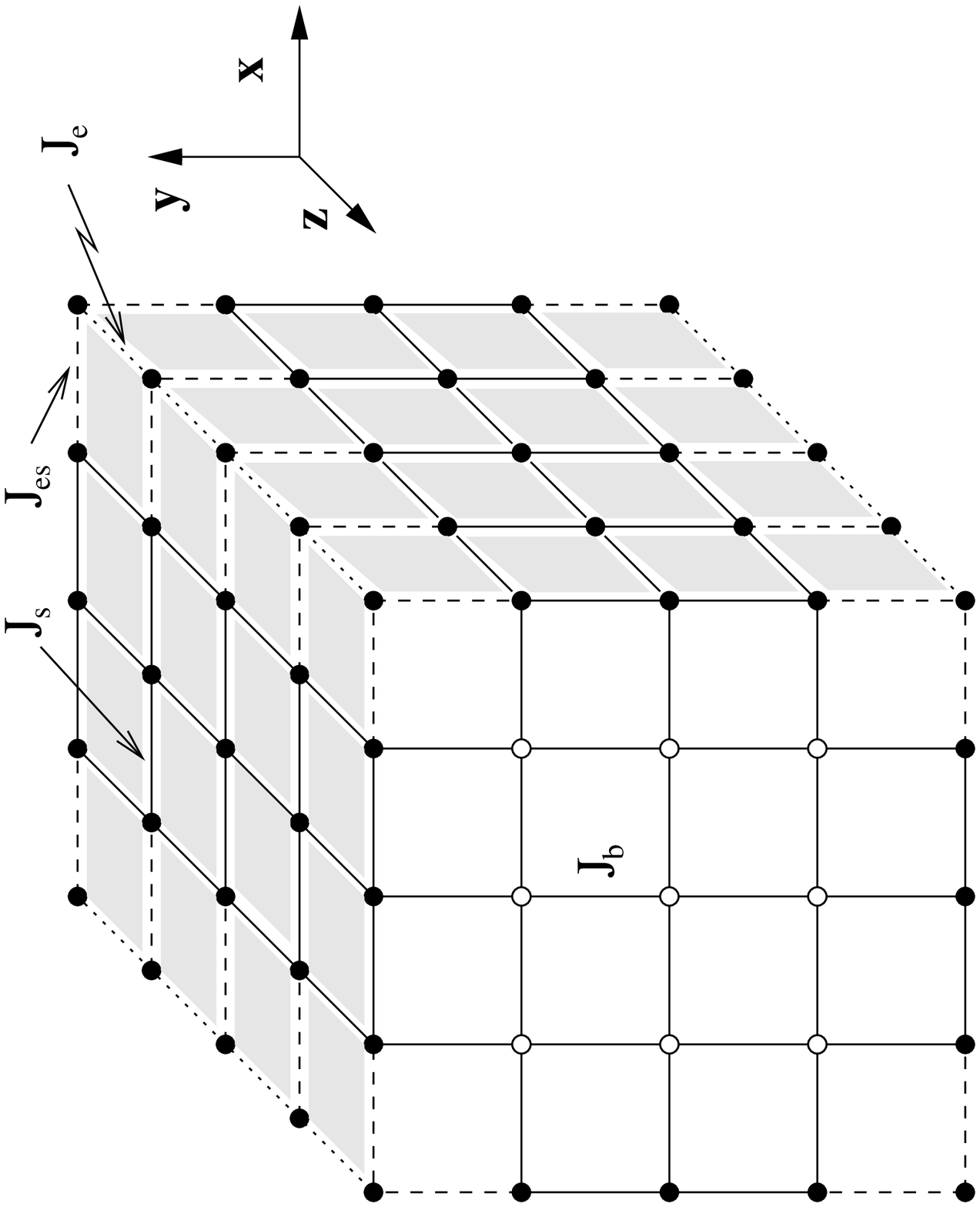,width=7.5cm,angle=270}}
\caption{Geometry of an Ising model with (100) and (010) surfaces forming edges
with opening angles $\theta = \pi/2$.}
\label{fig1} \end{figure}

\begin{figure}
\centerline{\psfig{figure=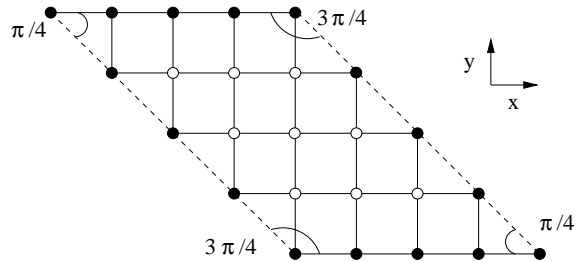,width=7.5cm,angle=270}}
\caption{Geometry of a model with edges with opening angles $\theta=\pi/4$ and
$\theta=3 \pi/4$. Shown is a cut through the crystal perpendicular to the edge direction. The full lines
represent the couplings between
neighbouring spins, whereas the (110) surfaces are indicated by the dashed lines.}
\label{fig2} \end{figure}

The results presented in the following have been obtained with the
efficient one-cluster-flip algorithm \cite{Wol89}. Systems with $L \times L \times
N$ sites were considered, with $L \times L$ being the number of sites in the planes
perpendicular to the $z$-axis, and $N$ being the number of sites along that axis.
$L$ ranged from 10 to 80 and $N$ from 10 to 640. For the investigation of corner
critical behaviour Ising cubes with up to $80^3$ spins were simulated.

As we aim at studying properties of systems with infinitely long edges, finite-size effects have
to be monitored closely. The Monte Carlo system should be large enough to reproduce
the thermodynamic values of the bulk and the surface magnetization 
sufficiently far away from the edge. In general, at a given temperature,
edge quantities must be stable against enlargening the system size. The same holds for corner 
quantities. 

An interesting quantity in systems without corners
is the magnetization per site for lines parallel to the $z$-axis:
\begin{equation}
m(x,y)= \langle | \sum\limits_z S_{xyz} | \rangle / N .
\end{equation}
The sum runs over the $N$ spins in a line, with $x$ and $y$ being fixed. The brackets denote
thermal averages. As usual, the absolute value is taken to avoid vanishing magnetizations
for finite systems.

From the line magnetizations, various magnetizations of interest are obtained. The edge
magnetization $m_2$ is identical to
\begin{equation} 
m_2 = \langle | \sum\limits_z S_{x_0y_0z} | \rangle / N
\end{equation}
where $(x_0y_0z)$ denotes an edge site. The surface magnetization results
from the line magnetization in the center of a surface, whereas the bulk magnetization
is given by the line magnetization in the middle of the crystal.

The local magnetization $m_l(x,y,z)$ may be obtained by computing the
correlation function $\left< S_{xyz} \, S_{x'y'z'} \right>$ between topologically
equivalent sites ($xyz$) and ($x'y'z'$) with maximal separation distance.
In this way, the corner magnetization of Ising cubes with linear dimension $L$ is given by
\begin{equation}
m_3 = \sqrt{ \langle S_{000} \, S_{LLL} \rangle }.
\end{equation}
I refer to corner quantities by the subscript 3 (indicating that three planes meet at this 
point), whereas edge quantities have the subscript 2.

\section{Ordinary transition}
In \cite{Car83}, Cardy considered $d$-dimensional systems with $(d-1)$-dimensional hyperplanes
meeting at an angle $\theta$. At the bulk critical point
local critical exponents changing continuously with
$\theta$ are already obtained in mean-field approximation. 
%The edge magnetization critical
%exponent is for example given by
%\begin{equation} \label{gl:4_1}
%\beta_2^{MF}= \frac{1}{2} + \frac{\pi}{2 \theta}.
%\end{equation}
Cardy showed that all edge critical exponents can be obtained by combining
bulk and surface critical exponents with a new angle dependent edge exponent.
He also computed the value of this
exponent in first order of an $\epsilon=d-4$ expansion. From the renormalization group point of view, 
the angle dependence has its origin in
the invariance of the edge under rescaling. This makes the opening angle to a marginal variable,
which may then lead to angle dependent local critical exponents.

Edge magnetizations computed by Monte Carlo simulations are plotted in Figure 3 for
four different opening angles:  $\theta= \pi/4$, $\pi/2$,
$3\pi/4$, and $\pi$ (correponding to the free surface), with all coupling 
constants set equal to $J_b$. Only data
presenting no noticeable finite-size dependences are displayed. 
This is achieved by studying systems with a large number of edge sites, the
edge length being the crucial quantity \cite{Ple98}.
%The geometry 
%of Figure 1 is used to obtain the data for $\theta= \pi/2$, whereas 
%simulating the geometry of Figure 2 yields the data for 
%$\theta= \pi/4$ and $\theta= 3\pi/4$.

\begin{figure}
\centerline{\psfig{figure=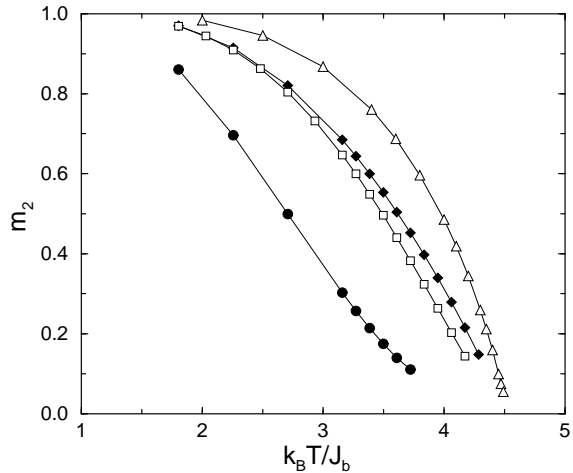,width=7.5cm,angle=270}}
\caption{Edge magnetization versus temperature 
for various opening angles $\theta$: $\pi/4$ (circles), $\pi/2$ (squares), $3 \pi/4$ (diamonds),
and $\pi$ (triangles). Only data not affected by finite-size effects are
displayed. Error bars are smaller than the size of the symbols.} 
%The largest system for the opening angle
%$\pi/2$ contained $60 \times 60 \times 640$ spins.}
\label{fig3} \end{figure}

At the ordinary transition, the edge magnetizations vanish at the bulk 
critical temperature $k_B \, T_c/J_b \approx 4.5115$ with a power law
behaviour $m_2 \sim (T_c - T)^{\beta_2}$. Here, $\beta_2$ is the critical exponent
of the edge magnetization. In order to estimate $\beta_2$, one may study \cite{Ple98a}
an effective temperature dependent exponent $\beta_{2,eff}(t)$ defined by
\begin{equation}
\beta_{2,eff}(t) = \mbox{d} \ln m_2 / \mbox{d} \ln t
\end{equation}
where $t= | T - T_c| /T_c$ is the reduced temperature. Certainly, on approach to $T_c$,
$t \longrightarrow 0$, $\beta_{2,eff}$ becomes the asymptotic exponent $\beta_2$.
%Because the simulations are performed at discrete
%temperatures, $t_i$, $\beta_{eff}$ is replaced by
%\begin{equation} 
%\beta_{eff}(t)= \ln (m_2(t_i)/m_2(t_{i+1}))/\ln (t_i/t_{i+1})
%\end{equation}
%with $t=(t_i+t_{i+1})/2$ (alternatively $t=\sqrt{t_i \, t_{i+1}}$). 
Very accurate Monte Carlo
data are required to get reliable estimates for the critical exponents.
 
\begin{table}
\caption
{Predictions for the edge magnetization critical exponents 
$\beta_2(\theta)$ from various methods. MF: mean-field
approximation, RNG: renormalization group theory, HTS: high temperature series expansions, MC:
Monte Carlo simulations.}
\vspace*{0.2cm}  
\begin{tabular}{|c||c|c|c|c|}
\hline
 & $\pi/4$ & $\pi/2$ & $3 \pi/4$ & $\pi$ \\
\hline\hline
MF \cite{Car83} & 2.50 & 1.50 & 1.17 & 1.00 \\
\hline
RNG \cite{Car83} & 2.48 & 1.39 & 1.02 & 0.84 \\
\hline
HTS \cite{Gut84} & 2.30 & 1.31 & 0.98 & 0.81 \\
\hline
MC    & $2.30 \pm 0.10$ & $1.28 \pm 0.04$ &$0.96 \pm 0.02$ & $0.80 \pm 0.01$ \\
\hline
\end{tabular}
\end{table}
%\begin{figure}
%\centerline{\psfig{figure=figure4.eps,width=7.5cm,angle=270}}
%\caption{Edge magnetization effective exponent as function of the reduced temperature $t$
%for various opening angles. The meaning of the symbols is the same as in Figure 3.
%The dashed lines
%represent the results obtained in an $\epsilon$-expansion \cite{Car83}, the dot-dashed lines
%indicate the values computed using high temperature series expansions \cite{Gut84}.}
%\label{fig4} \end{figure}

%Figure 4 shows the edge effective exponents derived from the edge magnetizations
%displayed in Figure 3. 
As the slope of $\beta_{2,eff}$ changes rather mildly at sufficiently
small reduced temperatures, meaningful estimates for the asymptotic exponent $\beta_2(\theta)$
are feasible. The values for $\beta_2(\theta)$ obtained in
this way \cite{Ple98} are listed in Table 1 and may be compared with results obtained
from mean-field theory \cite{Car83}, renormalizations group calculations to first order
in $\epsilon$ \cite{Car83}, or high temperature series expansions \cite{Gut84}.
The predictions of the renormalization group
seem to be systematically too large, as suggested both by high temperature series expansions
and our Monte Carlo simulations. The last two methods yield results which
are in close agreement with each other. 

Changing the values of the coupling constants or rotating the crystal about the $z$-axis
at fixed opening angle $\theta$ (the wedge is then formed by a different pair of surfaces)
does not alter the critical exponents \cite{Ple98}, in accordance with the expected
universality at the ordinary transition.

A similar analysis of the corner magnetization effective exponent yields for an Ising cube the
corner magnetization critical exponent $\beta_3 = 1.77 \pm 0.05$
\cite{Ple00}, significantly lower than the mean-field value $\beta_3^{MF}=2$ \cite{Igl93}.

\section{Surface transition}
When approaching the surface transition from high temperatures, 
the surface of a semi-infinite system orders whereas the bulk
remains disordered. In a three-dimensional system the critical fluctuations are then
essentially two-dimensional, and the surface critical exponents reflect the reduced dimensionality
by taking the values of the corresponding two-dimensional bulk system. At this transition,
an edge may be viewed as a local perturbation, acting
presumably like a line defect in a two-dimensional system.

Two-dimensional Ising models with defect lines have been shown to display non-universal
critical exponents \cite{Bar79}. 
A chain defect is formed by a column of perturbed couplings with strength $J_{ch}$,
whereas at a ladder defect modified couplings of strength $J_l$ connect spins
belonging to two neighbouring columns. In both cases 
local magnetization critical exponent depend on the strengths of these defect couplings \cite{Bar79}.
In analogy with these results, one expects at the surface transition, for a fixed opening
angle, intriguing non-universal edge and corner critical behaviour in three-dimensional
Ising models \cite{Ple99,Ple00}.

In the following, I discuss the influence of edges with opening angle $\theta=\pi/2$, see Figure 1,
on the local critical behaviour at the surface transition. The ratio $J_s/J_b= 2 > r_{sp}$
is kept fixed,
whereas the edge-edge, $J_e$, and the edge-surface, $J_{es}$, couplings are varied.
To determine edge critical behaviour accurately, the critical temperature of the
surface transition, $T_s$,
needs to be determined accurately. Using standard finite-size analysis, one obtains
$k_B \, T_s/J_b=4.9575 \pm 0.0075$ for $J_s=2J_b$ \cite{Ple99}.
Because edges are one-dimensional, and all couplings
in the models are of short range, edge quantities become singular at $T_s$.

The profile of the line magnetization at the surface
%, defined by (see Figure 1 and eq.\ (2))
%\begin{eqnarray}
%m_{s}(i)& = & \left( m(x,L) + m(x,1) \right. \nonumber \\
%& & \left. + m(1,y) + m(L,y) \right)/4
%\end{eqnarray}
%with $i=x=y=1, \ldots L$, 
reflects the influence of bulk spins close to the surface and edges
as well as the strength of the couplings near the edge, $J_e$ and $J_{es}$. Typical profiles
are shown in Figure 4 for $J_e=J_{es}=J_s=2J_b$. The edge magnetization is $m(1,L)$.
At low temperatures, $m(x,L)$ increases monotonically with the distance from the edge.
The magnetization of the surface spins, which are connected directly to an ordered bulk spin,
is enhanced compared to the edge magnetization. In contrast, on approach to $T_s$, the ordering
of the spins falls off quickly by going from the surface to the bulk, and the surface
magnetization is pulled down below the edge magnetization by the coupling to the disordered
bulk spins. Roughly at $T_c$, a non-monotonic behaviour in the profile shows up, with
a maximum close to the edge. It may be explained by the fact that surface spins next to
but on different sides of an edge are more strongly connected to each other, through the same
neighbouring bulk spin, than spins on the flat part of the surface with the separation distance
of two \cite{Ple99}.

\begin{figure}
\centerline{
\psfig{figure=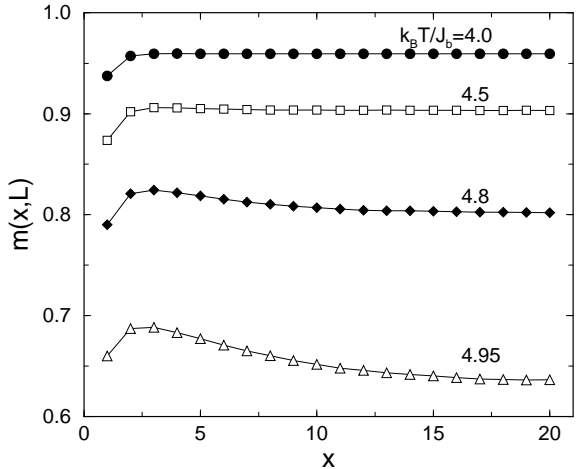,width=7.5cm,angle=270}}
\caption{Line magnetization profiles near an edge with opening angle $\theta=\pi/2$
at various temperatures. The couplings are $J_e=J_{es}=J_s=2J_b$. Systems with $40^3$ spins
have been simulated.}
\end{figure}

%Near the surface transition, 
%the edge acts like a defect line in an essentially two-dimensional bulk Ising model. The edge
%coupling $J_e$ then corresponds to a chain-like defect, the edge-surface coupling $J_{es}$
%to a ladder-type defect. The change in the topology at the edge compared to the surface amounts
%to a complicated ladder-type defect.

The effect on the edge critical behaviour of weakening
the interactions between the edge and the surfaces, $J_{es}$, is demonstrated in Figure 5.
A plateau-like behaviour is observed close to $T_s$, which facilitates
a precise determination of the asymptotic value. Obviously, the value of the critical exponent
is non-universal, changing continuously as a function of the coupling strength $J_{es}$, see
Table 2. A similar dependency is found when $J_{es}$ is kept fixed whereas $J_e$ is changed
\cite{Ple99}. The
findings of Figure 5 and Table 2 may be related to the 
reported results on two-dimensional Ising models with a ladder
defect \cite{Bar79}. The ladder column corresponds to the edge, and the ladder couplings $J_l$ reflect
not only the edge-surface interaction $J_{es}$ but also the reduced connectedness to bulk spins
at the edge compared to the surfaces. For $J_{es}=J_e=J_s$, the critical exponent of the edge
magnetization has the value $\beta_2= 0.095 \pm 0.005$, significantly lower than the
critical exponent of the perfect two-dimensional Ising model, $\beta = 1/8$. 
Of course, the non-monotonic profiles of the line magnetization
suggest that a closer analogy between edge properties and descriptions by two-dimensional
Ising models with defect lines would require more complicated, extended ladder-type defects.

\begin{figure}
\centerline{
\psfig{figure=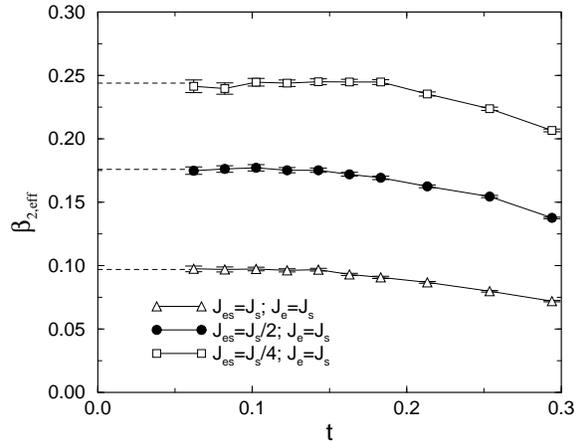,width=7.5cm,angle=270}
}
\caption
{Effective edge exponents as function of the reduced temperature 
for various values of the coupling $J_{es}$ with $J_e=J_s=2J_b$.
Only data not affected by finite-size effects are displayed. The largest system simulated
contains $80^3$ spins. Error bars result from averaging over several realizations.}
\end{figure}

Corner criticality in three-dimensional Ising models deserves to be analyzed at the surface
transition as well. As edges are local perturbations acting similar to defect lines in
two-dimensional models, the corners of a cube may be interpreted as intersection points of
three defect lines. 

As for the edge, the local magnetization profiles show a 
non-monotonic behaviour along paths from the edges or corners towards the center of the
surface, reflecting the influence of bulk spins. At lower temperatures, the profil
is monotonic, with the smallest magnetization at the corners. Close to $T_s$, a monotonic
behaviour is also expected, with the highest magnetization at the corners. Due to finite-size
effects, this last behaviour is not observed in the simulated finite cubes \cite{Ple00}.

The analysis of corner magnetization effective exponents yields again critical exponents
changing continuously with the strengths of the local couplings, see Table 2. Note that
corner criticality has not been studied for the case $J_{es}=J_s/4$. Changing corner critical
exponents are also observed when varying $J_e$ \cite{Ple00}. This non-universal corner criticality
is understood by the analogy to two-dimensional Ising models with star defects formed by three intersecting
ladder defects \cite{Hen88}. 

\begin{table}
\caption
{Edge and corner magnetization critical exponents $\beta_2$ und $\beta_3$
at the surface transition for systems with opening angles $\theta=\pi/2$ and
$J_e=J_s=2 J_b$.}
\vspace*{0.2cm}  
\begin{tabular}{|c||c|c|c|}
\hline
 & $J_{es}=J_s$ & $J_{es}=J_s/2$ & $J_{es}=J_s/4$ \\
\hline\hline
$\beta_2$ & $0.095 \pm 0.005$ & $0.176 \pm 0.005$ & $0.244 \pm 0.005$ \\
\hline
$\beta_3$ & $0.06 \pm 0.01$ & $0.26 \pm 0.02$ & \\
\hline
\end{tabular}
\end{table}

\section{Conclusion}
Critical phenomena may occur not only in the bulk but also at surfaces, edges, and corners.
I have presented results of an extensive Monte Carlo study of edge and corner criticality in three-dimensional
Ising systems. At the ordinary transition, edge and corner critical exponents have been computed
and compared to analytical predictions. Whereas the critical exponents depend on the wedge angle,
they do not depend on microscopic details as the strengths of the local couplings. This is different at
the surface transition where coupling dependent edge and corner critical exponents are observed. 
This non-universal behaviour is understood by analogy with exact results on two-dimensional Ising
models with defect lines.  

\ack
It is pleasure to thank Walter Selke for the fruitful and stimulating
collaboration on the edge and corner criticality.

\end{document}